# Shear-induced electrical changes in the base of thin layer-cloud

R. Giles Harrison[1], Graeme J. Marlton[1], Karen L. Aplin[2], Keri A. Nicoll[1,3]

[1] Department of Meteorology, University of Reading, UK

[2] Department of Physics, University of Oxford, UK, *now at* Department of Aerospace Engineering, University of Bristol, UK

[3] Department of Electronic and Electrical Engineering, University of Bath, UK

**Abstract**

Charging of upper and lower horizontal boundaries of extensive layer clouds results from current flow in the global electric circuit. Layer-cloud charge accumulation has previously been considered a solely electrostatic phenomenon, but it does not occur in isolation from meteorological processes, which can transport charge. Thin layer clouds provide special circumstances for investigating this dynamical charge transport, as disruption at the cloud-top may reach the cloud base, observable from the surface. Here, a thin (~300 m) persistent layer-cloud with base at 300 m and strong wind shear at cloud-top was observed to generate strongly correlated fluctuations in cloud base height, optical thickness and surface electric Potential Gradient (PG) beneath. PG changes are identified to precede the cloud base fluctuations by 2 minutes, consistent with shear-induced cloud-top electrical changes followed by cloud base changes. These observations demonstrate, for the first time, dynamically driven modification of charge within a layer-cloud. Even in weakly charged layer-clouds, redistribution of charge will modify local electric fields within the cloud and the collisional behaviour of interacting charged cloud droplets. Local field intensification may also explain previously observed electrostatic discharges in warm clouds.

*Keywords:* atmospheric electricity; stratiform cloud; Kelvin-Helmholtz billows; cloud microphysics;

1. **Introduction**

Stratiform clouds cover about 40% of the planet (Klein and Hartman, 1993), and have an important role in the radiation balance, which is strongly influenced by cloud microphysical properties such as the droplet size distribution. For horizontally extensive layer-clouds, such as those providing whole sky coverage as seen from a particular observing site, the cloud acquires electric charge at its upper and lower horizontal boundaries. Acquisition of charge at the cloud boundary results from current flowing in the global atmospheric electric circuit, through which charge transfer occurs between thunderstorm regions and distant regions of undisturbed weather. Charging of small droplets has been suggested to influence their behaviour through modifying collision (Tinsley *et al* 2000; Khain *et al* 2004) and activation (Harrison and Ambaum 2008) processes. Clear evidence for upper and lower edge electrification (of positive and negative charge respectively) has now been obtained at multiple sites using instrumented balloon soundings (Nicoll and Harrison, 2016), and charging of the lower boundary has also been found to be directly observable from surface sensors, if the cloud base is below 1500 m (Harrison et al, 2017a).

Previous work on layer-cloud charging has concentrated on a decoupled electrostatic representation, assuming the cloud is meteorologically passive (e.g. Zhou and Tinsley 2012). Observations (Nicoll and Harrison, 2016) show that, whilst the straightforward electrostatic description of upper edge positive charge and lower edge negative charge does emerge on average, individual layer-clouds show substantial variability with charge present throughout the vertical extent of the cloud as well as at the upper and lower edges. There are several possible reasons for this, at least two of which are meteorological in origin. Firstly, the cloud boundary properties depend on the local meteorological conditions, through the effect of the temperature inversion at the cloud-top and updraft strength at the cloud base. These influence the vertical electrical conductivity gradient at each horizontal cloud boundary and, in turn, the cloud edge charge (Nicoll and Harrison, 2016). Secondly, mixing and turbulence occur within layer-clouds (Shao et al, 1997), which transports charge from the edge region into the main body of the cloud. A purely electrostatic model can therefore only provide an approximate representation of layer-cloud electrification, particularly if based on simple geometry, as the charges generated at the cloud edges are always susceptible to dynamical transport and turbulent mixing processes.

In this paper, above-cloud instability is shown to influence the charged regions within a thin but persistent low-level extensive layer-cloud. Remote electrostatic sensing is used to investigate fluctuations in the charged cloud base and the effect of enforced motion disturbing charge at the upper cloud edge is explored.

2. **Meteorological circumstances**

(a) *Instrumentation*

The vertical electric Potential Gradient (PG) at the surface[1] is a widely observed property in atmospheric electricity, and shows appreciable variability on many timescales from minutes to days arising from weather, space weather and air pollution changes (Harrison and Nicoll, 2018). In the specific circumstances of persistent layer-clouds with a cloud base below about 1500 m, Harrison et al (2017a) demonstrated that cloud base charge influences the PG measured at the surface. As the cloud base lowers, the surface PG is reduced by the increasing proximity of negative charge in the cloud base. For these studies the surface PG was obtained using an upwards-facing field mill, and the cloud base height found using a laser ceilometer. At the Reading University Atmospheric Observatory[2] (RUAO) located at 51.44136°N, 0.93807°W, a Chubb Instruments all-weather field mill, JCI131 is used, sampled at 1s intervals, co-located with a Vaisala CL31 ceilometer. The ceilometer determines the cloud base height to a resolution of 9m by a time-of-flight measurement from upward-propagating pulses of near infra-red laser light, together with a profile of the backscatter of the laser light up to cloud base. Little backscatter information is available within the cloud above the lower cloud-air boundary, as the CL31's laser light is strongly attenuated when it reaches the cloud. The Reading CL31 is configured to sample every 5 s, but subsequent samples cannot be regarded as entirely independent of each other as some smoothing and processing is applied by the manufacturer's internal algorithms (Kotthaus et al, 2016). Averages at one minute intervals are constructed for the first part of the analysis here, to both improve the vertical resolution and sample independence, with the 5 s raw data considered further in the second part.

The RUAO also operates broadband radiation instruments to determine the solar radiation components at the surface (including $S_g$, the global solar radiation and $S_d$, the diffuse component), and the upwelling and downwelling longwave radiation ($L_u$ and $L_d$ respectively). Whilst 1s samples are available for the solar radiation, the response time of the radiometer instruments themselves is about 15s (Harrison, 2015). Further, during the period of the measurements discussed, the $L_d$ values were only available as 1 minute mean values.

---

[1] The standard convention used here is that the Potential Gradient is $-E_z$, where $E_z$ is the vertical component of the atmospheric electric field. The PG is positive in fair weather.
[2] http://www.met.reading.ac.uk/observatorymain/

In addition, two instrumented balloon soundings were made. These used enhanced RS92 radiosondes, each carrying a pair of solar radiation sensors (Harrison et al, 2016). The characteristic cloud-air transition in the solar radiation variability measured on a swinging radiosonde platform (Nicoll and Harrison, 2012) was used to determine the cloud-top position.

(b) *Conditions*

The layer-cloud investigated here persisted over RUAO during 19[th] March 2015 (year day 78 of 2015) and dissipated after local noon on 20[th] March 2015. The cloud received unusually close scrutiny because it obscured the partial solar eclipse which occurred on 20[th] March, and the associated atmospheric electricity and meteorological conditions are extensively described in Bennett (2016) and Burt (2016) respectively.

Figure 1 shows ceilometer, PG and sounding measurements made beneath the layer-cloud. In figure 1a, time series are given of the backscatter profile and the retrieved cloud base height, as determined by the ceilometer's internal processing algorithm which uses the backscatter transition at cloud base. During almost all of day 78, substantial and repeated fluctuations are evident in the cloud base height, whereas for day 79, there is much less variability by comparison. (Images of the cloud base on the two days are given in figure A1, to show the different appearance of the cloud base on the two days). Figure 1b presents the time series of PG beneath the cloud layer. It also shows markedly more variability on day 78 compared with day 79.

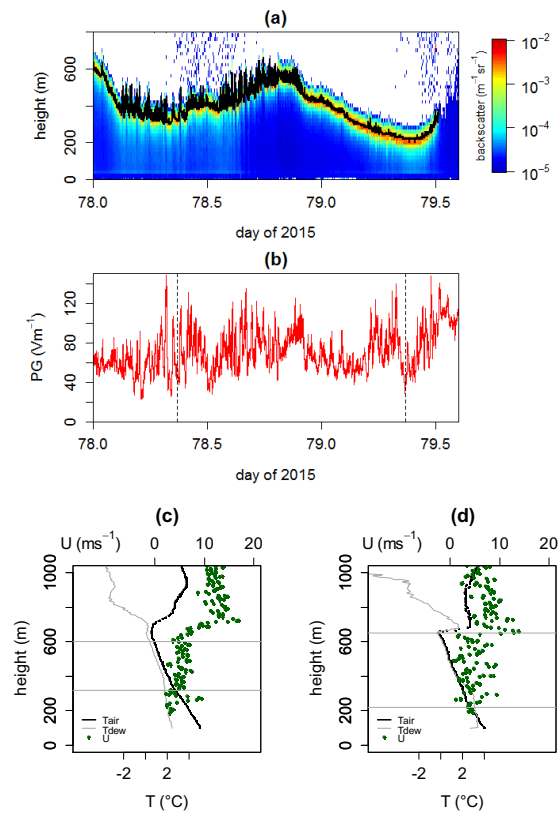

Figure 1. Time series of (a) backscatter obtained from a CL31 laser ceilometer above Reading on 19[th] and 20[th] March 2015 (year days 78 and 79), with the instrument-retrieved cloud base marked in black. (b) Time series of electric potential gradient (PG) measured at the surface. Dashed lines mark times of radiosonde launches. Sounding profiles obtained are shown in (c) and (d), of air temperature ($T_{air}$, black line) and dewpoint temperature ($T_{dew}$, grey line) and horizontal wind speed ($U$, green dots). The cloud base and top are marked with horizontal lines. (The cloud base is obtained from the ceilometer, and the cloud top from a solar radiation sensor carried on the radiosondes, with cloud top defined as the position where measured solar radiation variability is halfway between its in-cloud and clear air values).

Radiosondes were released from RUAO at 0845 UTC on both days, and data from the two soundings are shown in figures 1c and 1d and are summarised in Table 1. The soundings show thin cloud, of thickness 323 m on day 78 and 432 m on day 79. Temperature inversions define the top of the cloud, with the more marked inversion on day 79. Above the cloud at 800m, the relative humidity is 69% (day 78) and 79% (day 79). A strong difference between the two days is the wind shear at the cloud-top. For the day 78 sounding this exceeds 0.1 s$^{-1}$ ((m s$^{-1}$) m$^{-1}$) in the 120m above the cloud-top, whereas for the day 79 sounding it is much less, 0.006 s$^{-1}$. (Figures A1a and c provide skycam views taken close to the balloon release times).

**Table 1. Properties derived from the two atmospheric soundings**

| Day of 2015 | Cloud parameters | | | Cloud top gradients* | | Turbulence parameters | |
|---|---|---|---|---|---|---|---|
| | Ceilometer cloud base at launch (m) | Cloud top height (m) | Cloud depth (m) | Wind shear ((ms$^{-1}$)m$^{-1}$) | Temperature change (K m$^{-1}$) | Richardson number Ri | critical wavelength $\lambda_c$ (m) |
| 78 (19$^{th}$ March) | 318 | 641 | 323 | 0.11 | 0.015 | 0.04 | 38 |
| 79 (20$^{th}$ March) | 218 | 650 | 432 | 0.006 | 0.032 | 0.6 | 0.06 |

*calculated across 120m layer at cloud top

*(c) Turbulence*

Horizontal wind shear is well-known to produce instability (Richardson 1920), which is apparent in the generation of Kelvin-Helmholtz (K-H) waves made visible by cloud formation (Browning 1977). These or related turbulent sources of regular motion, may therefore be driving the fluctuations observed in the cloud base on day 78, which are not apparent on day 79 when the wind shear in the sounding is negligible. This possibility is supported from examination of the hourly values of wind speed and wind shear calculated by ECMWF (figures A2a and A2b), without assimilating the Reading radiosonde data. Figure A2b shows that the strongest wind shear occurs at the same time as the cloud base fluctuations. Direct comparison of the forecast with the radiosonde suggests that the ECMWF model was better at predicting the mean horizontal wind speed than the wind shear and therefore the exact location of the shear generating region, which is likely to be highly local, can only be regarded as approximate.

The presence or absence of turbulence at the cloud-top can be assessed using the Richardson number Ri (Richardson, 1920; Miles, 1961), found from the vertical gradients of potential temperature *θ* and horizontal wind speed components *u* and *v* as

$$Ri = \frac{g}{\theta} \frac{\frac{d\theta}{dz}}{\left(\frac{du}{dz}\right)^2 + \left(\frac{dv}{dz}\right)^2} \quad (1),$$

where *z* is the height coordinate and *g* the gravitational acceleration. Turbulence is conventionally regarded as present when Ri<0.25 (Miles 1961), although in some circumstances this threshold may be larger (Zilitinkevich et al, 2008; Baklanov et al, 2011). Evaluating the vertical change in wind speed and temperature over the transition distance of 120m evident at the cloud-top in figure 1c for both days, Table 1 shows that Ri indicates turbulence at cloud-top on day 78 but not on day 79. Further evidence that the variations in cloud base are associated with turbulence is provided from the power spectrum of the cloud base height time samples (figure A2c), which also shows the Kolmogorov -5/3 spectral slope typical of a turbulent flow (Kolmogorov 1941).

For the case of K-H instability, which generates internal breaking waves from gravity waves (gravity-restored displacements), a critical wavelength $\lambda_c$ is identified by

$$\lambda_c = \frac{\pi \rho (\Delta U)^2}{g \Delta T} \quad (2)$$

where the wind speed difference $\Delta U$ and temperature $\Delta T$ are evaluated cross the shear region and $\rho$ is the air density (Cushman-Roison, 2014). Only gravity waves with wavelengths shorter than $\lambda_c$ grow into K-H billows. Table 1 also provides $\lambda_c$ evaluated from equation (2) for both days using the sounding information, from which it is apparent that K-H oscillations with wavelengths of tens of metres are indicated to be possible on day 78.

The fluctuations in the cloud base apparent on day 78 can therefore be attributed to turbulent motions generated by shear instability above the cloud top.

### 3. Observations of cloud fluctuations

The electrical variations associated with the dynamical instabilities are now considered further, firstly for the slow periodic variations observed of the order of 10-20 minutes, and secondly for the rapid steps that occurred on timescales of 1-2 minutes.

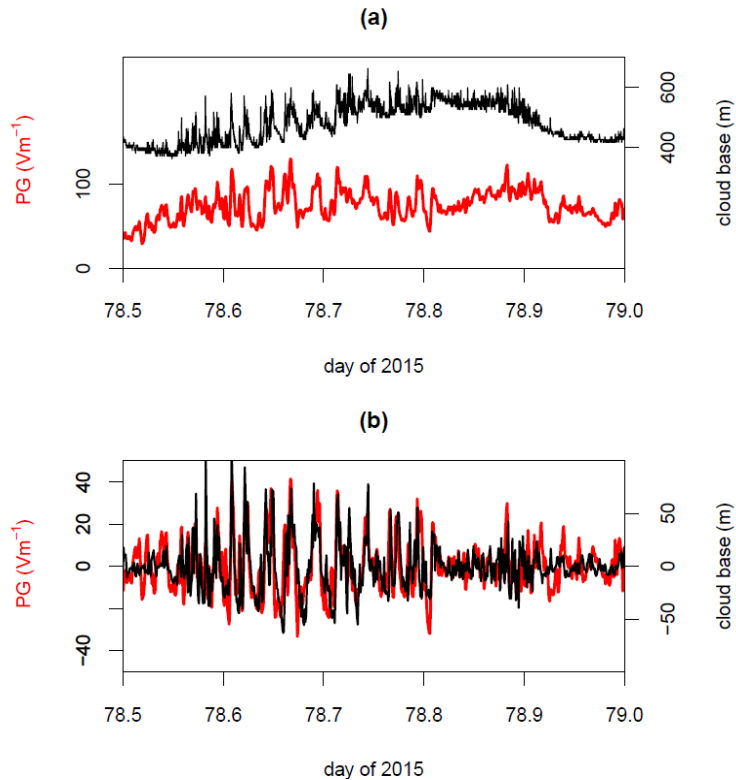

Figure 2. Time series of cloud base height (thin black line) and surface Potential Gradient (PG, thick red line), as (a) 1 minute mean values and (b) high pass filtered 1 minute values, with variations slower than 20 minutes removed.

(a) *Slow symmetric fluctuations*

Figure 2a shows the detail of cloud base height and surface Potential Gradient (PG), from which it is apparent that both quantities are well correlated. These slow fluctuations can be extracted by filtering, to remove variations with periods greater than 20 minutes. Figure 2b shows the same data after high pass filtering from which the close, quasi-oscillatory relationship between the cloud base height and PG is immediately evident. It is highly unusual to be able to identify the origin of surface PG variability so explicitly, because multiple sources of variability are usually present. For the mean cloud height in this case (483 m), cloud base height fluctuations of up to about ±70 m are associated with PG fluctuations of ±30 V m$^{-1}$. This sensitivity is larger than the typical 0.1 (Vm$^{-1}$) m$^{-1}$ found previously for slow cloud base changes (Harrison et al 2017a,b), suggesting that the charge varying in this case is several times greater.

Closer examination of the time series in figure 2b, however, indicates that the PG changes often occur before the cloud base changes, i.e. that there is a lagged response. This adds weight to the possibility that the responses observed are not solely due to cloud base fluctuations, instead, for example, arising from vertical motion driven by the horizontal rolls above cloud top, combined with local turbulence. This motivates further investigation of the lag. In figure 3, composites are formed on the PG minima and maxima of the high pass filtered data to draw out the phasing of the PG and cloud base changes. Two periods are chosen for this, the second quarter of day 78 (i.e. 78.25 to 78.5) during which the radiosonde measurements of figure 1 were obtained and therefore when the shear is observed to be present, and the second half of the day (i.e. 78.5 to 79), during which there were larger fluctuations and the shear is inferred to continue from the ECMWF analysis. For both the PG minima and maxima, statistically significant minima and maxima in the cloud base follow two minutes later, with an in-phase response with the PG. This analysis verifies that the electrical changes occur before the cloud base change observed by the ceilometer. The correlation demonstrated in figure 2b is therefore not solely a result of vertical fluctuations in the position of the charged cloud base, as these would cause an immediate response in the surface PG beneath.

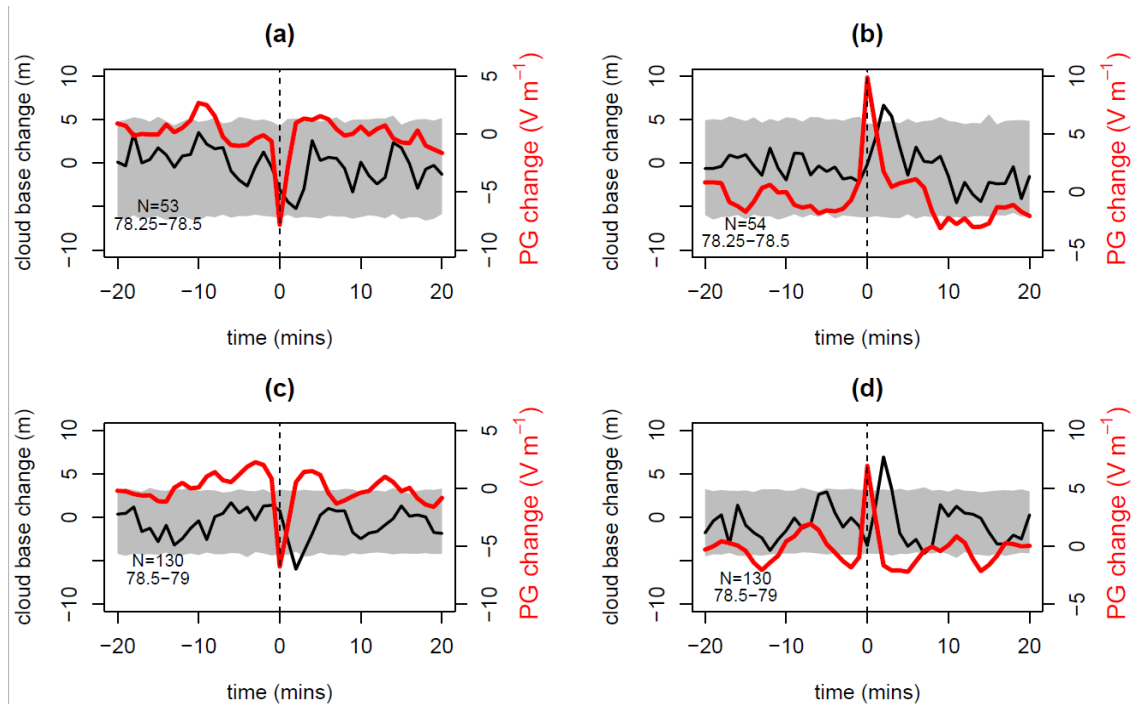

Figure 3. Averages of cloud base height fluctuations (black lines), calculated (composited) during the second quarter (a and b) and second half (c and d) of day 78, in both case on minima (a and c) and maxima (b and d) in the PG, using 1 minute high pass filtered values. The associated averages of changes in the PG values are also shown (red lines). Grey bands show the 95% confidence range on the cloud base values, from repeated sampling of the same number of events, but not associated with PG maxima and minima.

Some further insights into the cloud properties during the large cloud base height fluctuations can be obtained by examining the solar radiation measurements beneath the cloud. Figure 4a and 4b present cloud base and PG fluctuations, together with simultaneous co-located diffuse solar irradiance ($S_d$) measurements. Normalising the solar irradiance by the calculated (e.g. Harrison, 2015) top of atmosphere solar irradiance ($S_{TOA}$) at the same time and high pass filtering, figure 4c shows there is consistent behaviour in all three quantities during the latter half of the period considered when there are large fluctuations. The good correlation between cloud base height and solar radiation (shown in Table 2) for this period indicates that, when the cloud base rises and the PG increases, the optical thickness of the cloud is reduced, i.e. considered overall, the cloud thins significantly.

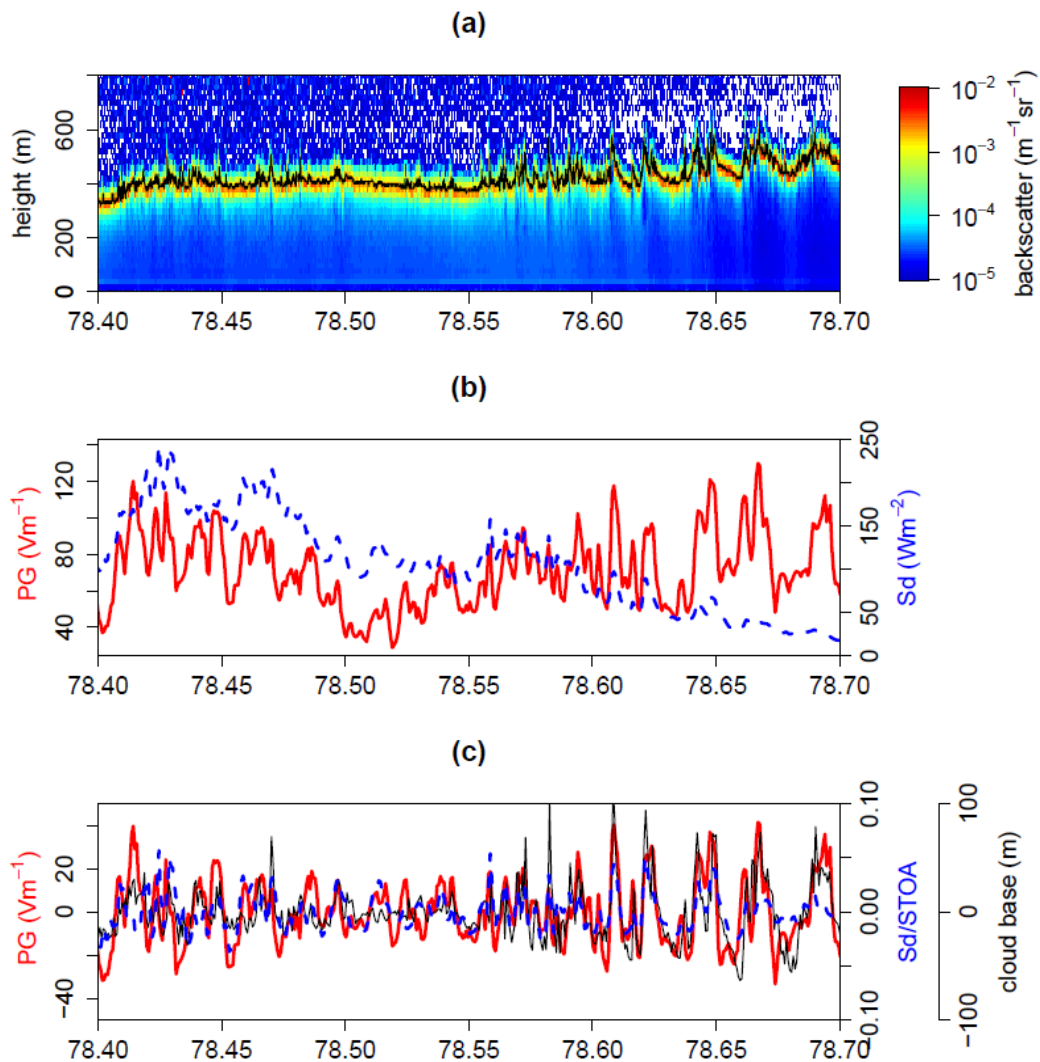

Figure 4. Selected time series of (a) backscatter and cloud base (black line), (b) PG (red line) and diffuse solar radiation ($S_d$, blue dashed line). (c) Time series of high pass filtered time series of cloud base (black line), solar radiation normalised by calculated top of atmosphere value at the same time ($S_d/S_{TOA}$, blue dashed line) and cloud base (black line). (Data are 1min averages from 1s samples.)

A similar timescale lag between changes in PG and radiation beneath a thin layer-cloud was found by Harrison and Ambaum (2009). These new direct observations of changes at cloud base remove the possible ambiguity of whether the previously reported effect arose from the cloud itself or through modification to the radiation environment below it. Further, the similar lag time found both for radiative instruments having a wide field of view and the ceilometer with a narrow point measurement indicates that the lag does not originate from horizontal propagation of cloud base anomalies which could affect the latter but not the former. Another physical length scale separating the two changes is therefore implied, to allow time for propagation of a structure, for example generated by a K-H wave, between the cloud top and bottom.

**Table 2. Pearson correlations between filtered variables from figure 4(c).**

| *Variables* | *Day fraction* 78.4 to 78.55 | 78.55 to 78.7 |
|---|---|---|
| Cloud base height and ($S_d/S_{TOA}$) | 0.02 | 0.37 |
| PG and ($S_d/S_{TOA}$) | 0.37 | 0.71 |

(b) *Rapid asymmetric fluctuations*

Further inspection of the cloud base height changes in figure 2b shows a range of amplitudes and shapes, and not solely slow undulations as discussed above in section 3(a). To examine the relationship between the electrical and cloud base changes more fully, all the rapid cloud base changes from day 78 have been plotted against the instantaneous PG at the time of the cloud change (figure 5), using the raw ceilometer data at 5s resolution. There are very many small cloud base changes which provide the central region of data in figure 5a, associated with statistical fluctuations between successive measurements, but there are also rather larger cloud base changes, which are much rarer: these have been highlighted around the edges of the data in figure 5a. It is clear that the distribution of these largest changes is asymmetric, in favour of positive (i.e. upwards) cloud base changes. Further, using boxplots to group the PG values associated with different rapid cloud base changes (figure 5b), both negative and positive rapid cloud base changes with magnitudes greater than 30m can be seen to be associated with an increased PG, compared with the small cloud base changes.

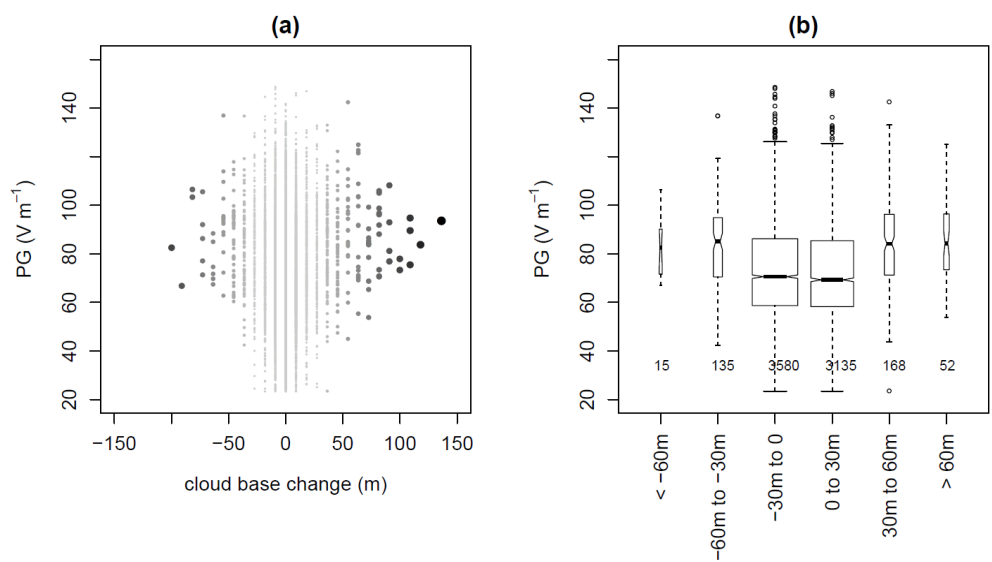

Figure 5. Cloud base changes on day 78 plotted against the instantaneous PG at the time of the change (ceilometer resolution 9m, 5s samples). (a) All cases, with the extreme values emphasised by increasing the size of the plotted point and its grayscale density in proportion to the cloud base change. (b) Cloud base changes from (a) binned into steps of 0 to ±30m, ±30 to ±60m and > ±60m, shown as boxplots with the number of cases marked.

Figure 6 examines the detail of variations associated with typical rapid cloud base displacement downwards (left hand columns) and upwards (right hand columns), again using the raw ceilometer data at 5s resolution. In both cases shown the PG increases during the upward cloud base displacement, as indicated by figure 5, but there are also associated changes in measurements of the downwards longwave radiation (a decrease) and shortwave radiation (an increase). In the case of the rapid upwards displacement, the upward fluctuations in cloud base recorded by the ceilometer are associated with a reduction in longwave radiation due to cooling, and an increase in shortwave radiation due to reduced optical thickness. The rapid change of cloud base position between consecutive 5s samples is, in some ways, reminiscent of the step change in electric field seen for a nearby lightning discharge.

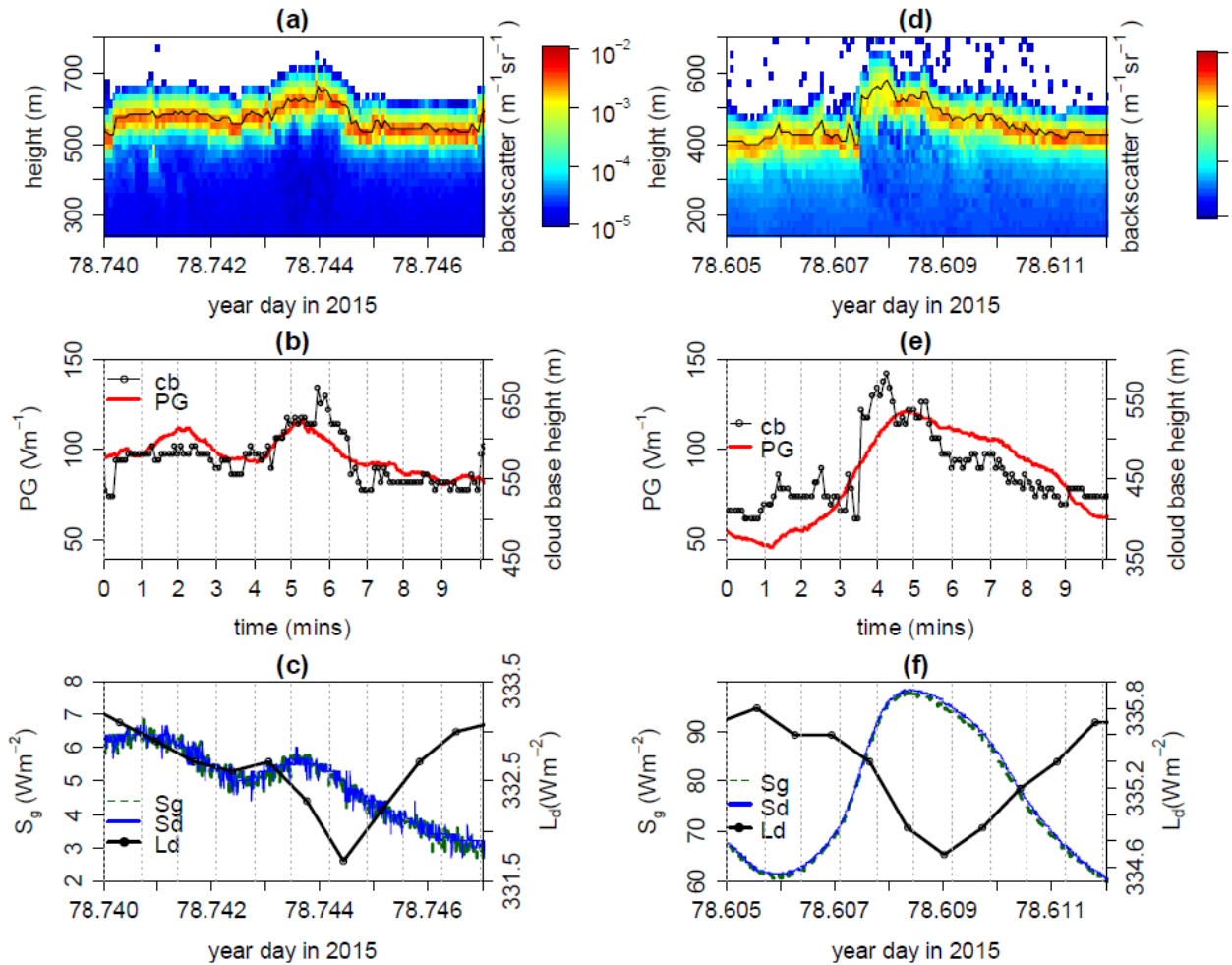

Figure 6. Changes associated with instantaneous fluctuation in the cloud base, downwards (left column) and upwards (right column). (a) and (d) show backscatter profiles and cloud base position (black line with points), (b) and (e) cloud base (black line with points) and PG (red line), (c) and (f), diffuse and global solar radiation ($S_g$ and $S_d$), and downwards long wave radiation ($L_d$). The ceilometer provides 5s data and the PG, $S_d$, $S_g$ are 1s values, $L_d$ are 1 min values.

Figure 7 averages together (composites) all the rapid large (> 30m) upwards and downward displacements in the raw 5s data of cloud base height, i.e. at 5 to 10 ms$^{-1}$. The composites produce the best summary of such changes, as they average many events together. These are for the same period as that shown in figure 4, in which there are 51 downward displacements and 69 upwards displacements. (This choice of step size is made so that it is several times greater than the minimum ceilometer resolution of 9m, likely to arise from statistical fluctuations between successive measurements.)

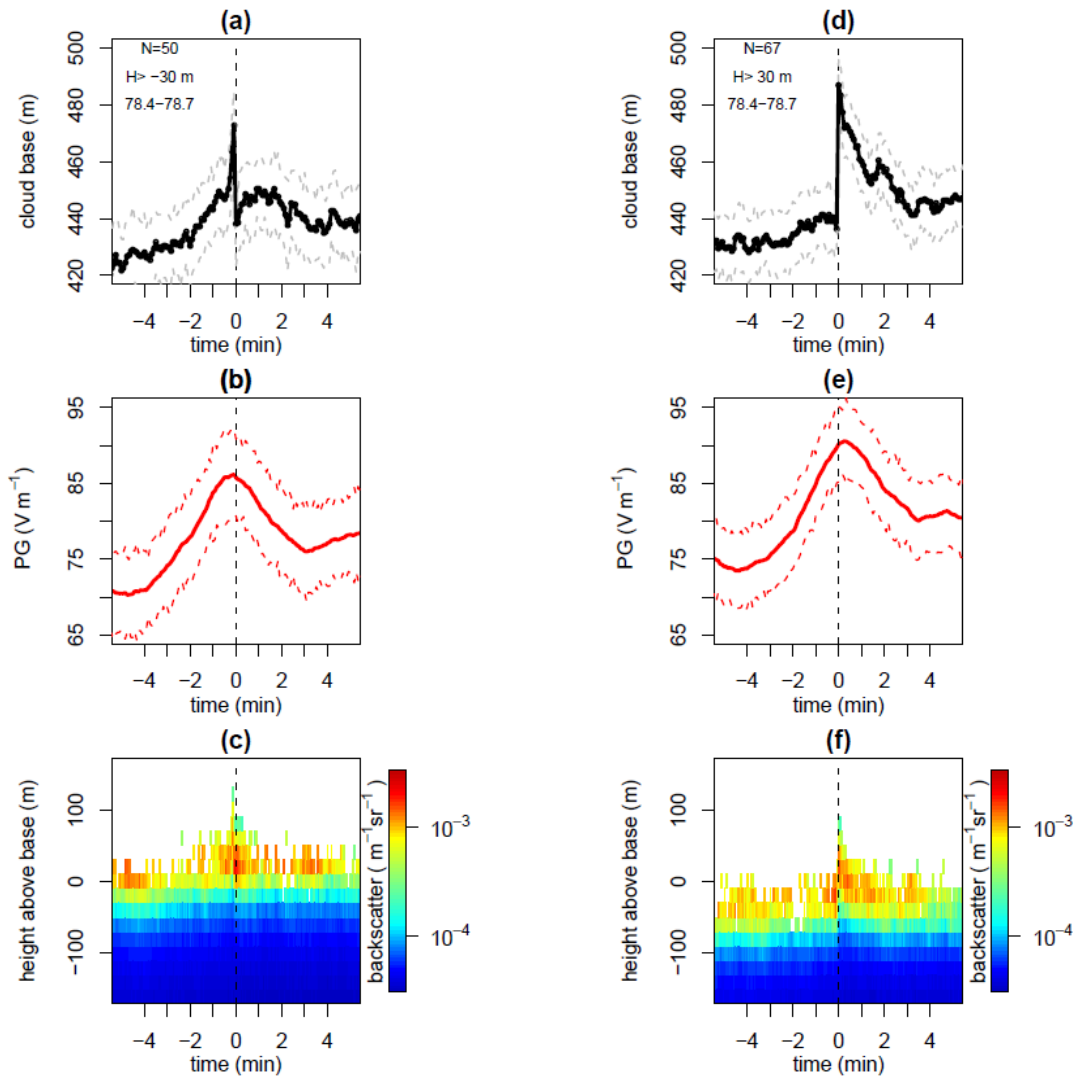

Figure 7. Composites between days 78.4 and 78.7 of variation in mean vertical position of cloud base using the instantaneous data ( (a) and (d) ), ( (b) and (e) )mean surface potential gradient, and ( (c) and (f) )median backscatter, reckoned from cloud base height at the event time. For (a)-(d) the 95% confidence on the line is marked. Left-hand panels are for rapid cloud base height decreases, and right-hand panels are for cloud base increases, with time axes all in minutes.

In figure 7(a), the rapid downward displacement of cloud base at $t$=0 in was preceded by a steady upwards drift in cloud base height. This is a different behaviour from the rapid displacements at $t$=0 in figure 7(d), in which there is effectively an isolated step change. In both the upward and downward cases, however, the averaged PG increased steadily before the event, as seen in figure 5. Notably, the PG reaches its maximum just before the cloud base decrease in figure 7a, and just after the cloud base increase in figure 7d. The rapid upward and downward displacements are therefore both generally associated with a prior increase in PG, and may be a consequence of the same disturbance event. The composited backscatter values in (c) and (f) both show an increase in backscatter at the displacement event time, with, necessarily, height variations before and after similar to those of the cloud base ((a) and (d)) respectively. Consequently, the rapid upwards cloud step is associated with displacing the dense backscatter region upwards.

These additional composites in figure 7 indicate two aspects that need explanation. Firstly, the, rapidity of the cloud steps, and secondly the PG increase for many minutes beforehand. The rapidity is not what would be expected from a wavelike undulation in the cloud base, which would cause a slow oscillation. Further, a simple propagating gap in the cloud base would lead to a PG change well correlated with the cloud base change. However, in the composites, the shape of the onset and recovery of the cloud base and PG changes are different.

4. **Discussion**

Previous work has highlighted the close relationship between surface atmospheric electrical changes and changes in the base of layer clouds, but in the observations reported here, electrical changes are detected before a cloud base height change. In the case of the slow fluctuations, the observed time lag is likely to be due to turbulence-induced downward motion of charge, based on the conclusions of section 2 that cloud-top or above-cloud turbulence is generated during day 78, and of section 3a that horizontal transport was unlikely. The presence of cloud-top charge, which in general is usually greater than cloud base charge (Nicoll and Harrison, 2016), is indicated by the larger PG sensitivity to cloud base changes than previously observed. Overall, the scenario envisaged to explain the slow symmetric variations with a lagged response is a dynamical disruption to the cloud-top charge which is sensed immediately through an induced PG fluctuation at the surface, and that the effects of the same disruption propagate to physically affect the cloud base some minutes later.

Beyond cloud-top disruption, if a region of positive charge were transported downwards by a wave structure or billow, and there is no change in the cloud base charge, the surface PG will increase with time as the charge descends. This may be apparent in figure 7b and 7d, in which a steady increase in PG begins about 4 minutes before the cloud step occurs. With a descent speed $W$ of 0.5 ms$^{-1}$, 120 m would be travelled in this time, or about half of the cloud depth. For the charge to be retained during its descent, the electrical conductivity must be sufficiently small for no appreciable dissipation of the charge to occur. In air of conductivity $\sigma$, the relaxation timescale controlling its discharge is $\varepsilon_0/\sigma$, where $\varepsilon_0$ is the permittivity of free space. Hence, if the charge descends at a speed $W$ through a cloud of conductivity $\sigma_{cloud}$, the time taken to pass through the cloud vertically must be shorter than the relaxation time scale, for the charge to be maintained. This provides a solely electrical condition on the maximum cloud depth $D$, as

$$D \ll \frac{W\varepsilon_0}{\sigma_{cloud}} \quad (3)$$

assuming that there is no additional contribution to the discharge process from turbulent mixing. The in-cloud conductivity is poorly known, but assuming $\sigma_{cloud}$~ 2 fSm$^{-1}$ (i.e. ~ 1/5th of the typical clear air conductivity at the surface) and $W$=0.5 ms$^{-1}$, equation (3) requires that $D$ should be less than 2212m, a condition which Table 1 indicates is easily fulfilled for the cloud circumstances described and therefore that charge would be sustained during vertical transport.

More extensive evidence demonstrating downward motion in a similar thin layer-cloud on the same day is available in data from Chilbolton, 55 km from Reading, where a ceilometer and an upwards-facing Doppler cloud radar were both operating on day 78 of 2015. By compositing the Doppler radar data on the upwards cloud steps found from the Chilbolton ceilometer (figure A3), downward motion (coloured blue, of about 0.2 to 0.5 m s$^{-1}$ beginning about 300m above the cloud base) becomes apparent within the cloud before the cloud base step. This begins slightly before an upwards cloud step, and is present through much of the vertical extent of the cloud after the step, which, for the thin cloud during these conditions, is likely to include the upper charge region. (The variability evident in the composite at about 400m above the cloud base is associated with the cloud-top).

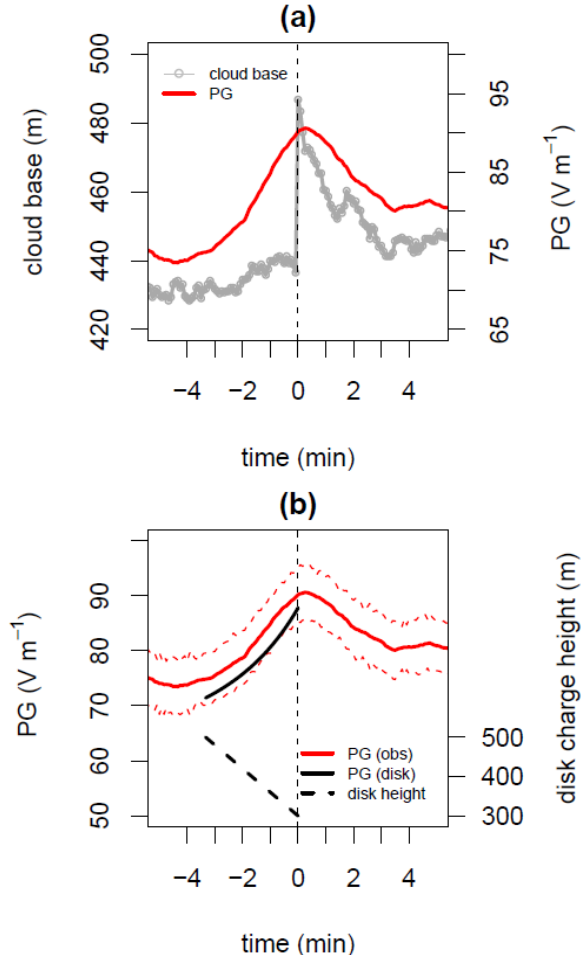

Figure 8. (a) Overlaid composites of cloud base (grey line and points, left-hand axis) and surface PG (red line, right-hand axis), from figs 7d and 7e. (b) Composited surface PG observations from (a), (solid red line, with 95% confidence limits dotted). The calculated surface PG is also included (black solid line), found from assuming a horizontal charged disk of radius 200 m carrying a charge density of +3 nC m$^{-2}$, descending at 1 ms$^{-1}$ from 500 m to 300 m, in a background surface PG of 60 Vm$^{-1}$. The variation of the disk charge position with time is given by the black dashed line (right-hand axis).

With this enhanced perspective from the Chilbolton data, the PG changes before an instantaneous cloud step are now re-visited (figure 8a). This shows that the PG changes are not strongly correlated with the cloud base before the cloud step, but afterwards they are more closely correlated. The pre-step increase in PG is therefore not strongly associated with the cloud base. Downward propagation of cloud-top charge, as observed at Chilbolton, is now considered as an explanation for the observed pre-step PG increase. Electrostatic representation of layer clouds by simple models such as parallel plate systems or point charges can only be regarded as approximate, but a disk charge model has previously proved useful to represent cloud base charge, using charge densities typically ~ -1 nC m$^{-2}$ on a disk of radius ~300 m (Harrison et al, 2017a). An assembly of positively charged droplets moving from the cloud top has therefore been considered as a migrating disk charge. For a charged disk of radius $R$, the electric field $E$ at a distance $H$ is derived from Gauss' Theorem as

$$E = E_0 + \frac{Q}{2\pi\varepsilon_0}\left[1 - \frac{H}{(H^2+R^2)^{1/2}}\right] \quad (4),$$

where $Q$ is the charge per unit area in the disk and $E_0$ is the background field (Jackson, 1962). If $H$ is the height of the disk charge, the electric field (or PG as $-E$) can be calculated beneath. Figure 8b shows the calculated variation in the PG at the surface, for a +3 nCm$^{-2}$ disk charge of radius 200 m descending from 500 m to 300 m altitude at 1 ms$^{-1}$, as informed by figure A3. A background PG of 60 Vm$^{-1}$ is assumed in the absence of the disk charge, to represent the fair weather PG, itself likely to be slightly suppressed by the presence of negative cloud base charge. Figure 8b shows agreement between the averaged and calculated PG change, with a non-linear increase in surface PG associated with the steady descent of the disk's positive charge. This indicates that the charge density and dimensions assumed, following Harrison et al (2017a), are not unreasonable, and that the observed PG increase before the cloud step is not inconsistent with a downward motion cloud top charge.

5. **Conclusions**

A close electrical association between the cloud base charge in low-level extensive layer-clouds and the surface PG has previously been established, using diurnal variations in cloud base height. Here, cloud base variations are examined within a persistent layer-cloud which is sufficiently thin for shear in the cloud-top to affect the cloud base. Observed surface electrical fluctuations are deduced to be caused by instability at or above the cloud-top, generating a downwards-propagating disturbance which ultimately reaches the cloud base, minutes later. Such a descent of charge from above can provide a quantitatively reasonable physical explanation for the steady increase in PG observed prior to more rapid cloud base changes.

This work demonstrates that atmospheric electrical properties are coupled with dynamical changes within layer-clouds, rather than a constant electrostatic system. This, in principle offers a possible method of remote sensing of cloud-top changes from the surface. It also illustrates, as the charge transferred by the dynamical transport is carried on droplets, that regions of oppositely charged droplets can be generated, locally modifying in-cloud electric fields. The interactions between charged drops have previously been demonstrated to differ from those of neutral drops, for example enhancing collision efficiencies and the timescale to produce rain. This may influence the cloud lifetime, and therefore the break-up of the layer cloud, which is an important climate parameter (Schneider et al, 2019). Further, as one region of charge is brought close to another of opposite polarity, the possibility exists that intense local electric fields may be generated, ultimately creating an electric discharge and generating radio frequency energy. Unexplained radio frequency emissions have previously been reported from warm stratiform clouds (Sartor, 1964) and drizzle producing clouds (Penzias and Wilson, 1970), for which the dynamically-forced transport of opposite charges towards each other, causing a discharge, provides a possible mechanism.

**Acknowledgements**

K.A.N. acknowledges NERC support through an Independent Research Fellowship (NE/L011514/1 and NE/L011514/2). The Copernicus Radar data and ceilometer data at Chilbolton used in figure A3 was provided by Chris Westbrook. The ECMWF forecast model data was obtained from the ECMWF MARS archive. Ken Bignell provided valuable information about the static discharges associated with non-thunderstorm clouds. The original data used is available from the corresponding author.

**Appendix**

Visual appearance of the cloud base can provide information on structure within a cloud and characteristic features can sometimes be repeatedly identified (e.g. Harrison et al, 2017c). Sky images are provided here for the days of interest from the Reading University Atmospheric Observatory, captured using an AXIS Q6035 Dome Network Camera looking in a northward-pointing direction. Figure A1 shows a series of images captured on 2015 day 78 ( (a) and (b) and day 79 ( (c) and (d) ). More structure, although not strongly developed, is apparent on day 78 than on 79.

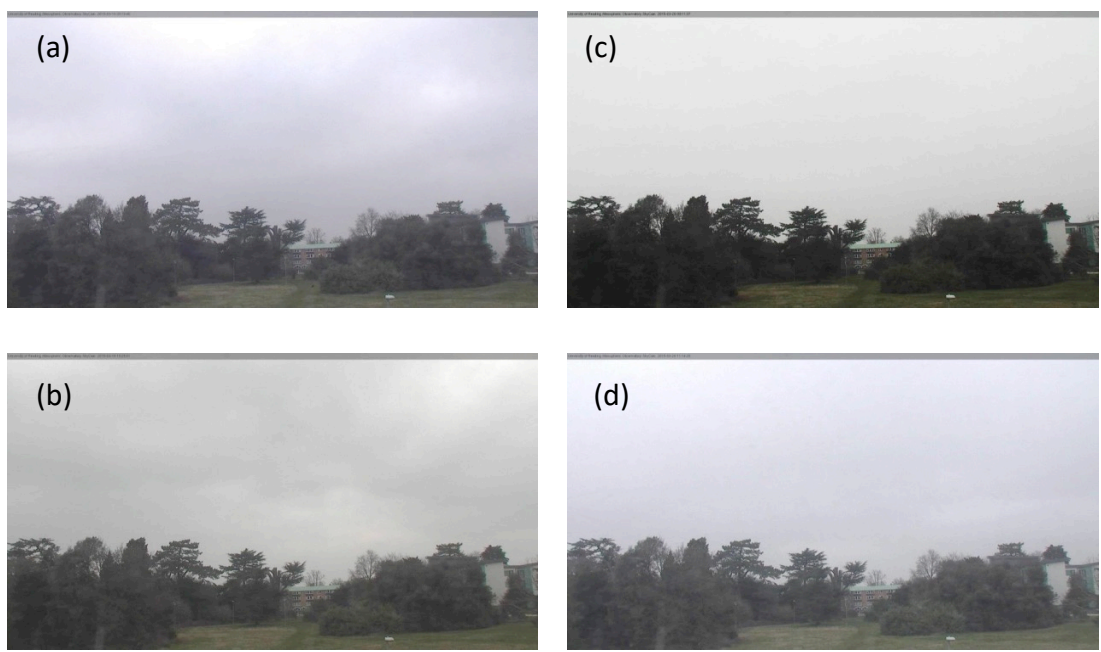

Figure A1. Skycam views northwards from the Reading University Atmospheric Observatory, on 19th March 2015 at (a) 0910 and (b) 1525, and 20th March 2015 at (c) 0911 and 1114.

Additional information about the state of the lower atmosphere over Reading during 2015 days 78 and 79 is provided in figure A2, from the ECMWF high resolution forecast model. (a) shows the ECMWF model output of the mean horizontal wind speed, (b) the time evolution of the wind shear above Reading and (c) the relative spectral power density in the Reading cloud base observations. A line showing a -5/3 gradient of the spectral power against frequency, characteristic of turbulence, is included.

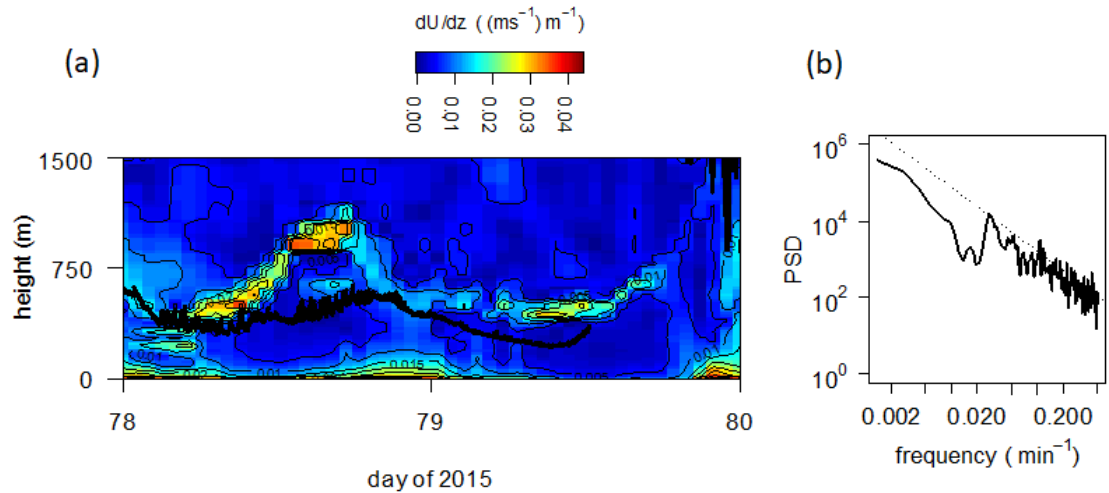

Figure A2. (a) and (b) time-height plots from the ECMWF high resolution forecast model. These are for the Reading grid square at 1 hour time steps, between the beginning of day 78 and the end of day 79, using forecasts initiated at midday and midnight. (a) mean horizontal wind speed ($U$) and (b) vertical wind shear ($dU/dz$), with the ceilometer cloud base measurements from Reading added to (a) (black line). (c) Relative power spectral density (PSD) calculated from the high pass filtered 1 minute cloud base height measurements, for the period of the cloud base fluctuations in day 78 (78.25 to 79). The dashed line marks a spectral slope of −5/3.

The properties of the thin cloud on 2015 day 78 were also studied at Chilbolton, Hampshire, 55 km from Reading were investigated using the Vaisala CL51 ceilometer sited there, which operates in the same way as the CL31 device at Reading. At Chilbolton there is also an upward-pointing Doppler radar (Copernicus) able to determine the speed of the cloud particles upwards or downwards. Figure A3 shows an average of the cloud particle speeds, around the times of rapid upward steps in the cloud base as determined by the ceilometer. The cloud particle speeds are shown spatially, with respect to the cloud base position found by the ceilometer. The variability in the upper part of the plot is associated with the cloud top. At the time of the cloud step, the Doppler radar shows that there is descending air within the cloud.

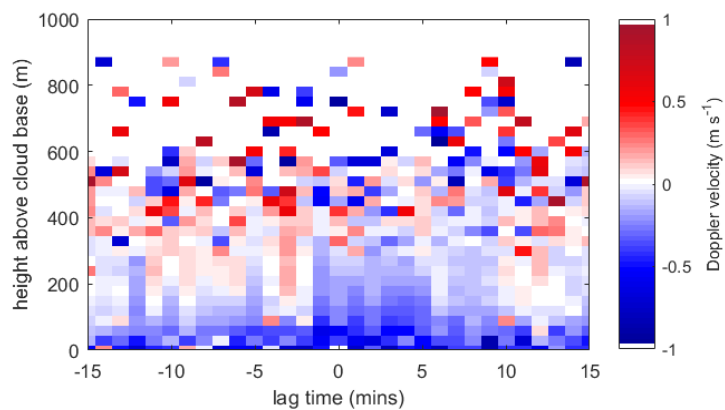

Figure A3. Analysis of layer cloud properties at Chilbolton during day 78 of 2015, by combining data from the site's laser ceilometer and cloud radar. The plot shows the averaged Doppler radar velocity within the cloud at a ceilometer upwards step, composited from 12 upwards cloud fluctuations exceeding 35m in the ceilometer data. The changes are found for the first 29 radar range gates above the mean cloud base height as found by the ceilometer, ± 15 minutes across each >35m step. (Blue colours show vertically downward wind directions.)